\documentclass[%
 reprint,
groupedaddress,
showpacs,preprintnumbers,
 amsmath,amssymb,
aps,
prd,
]{revtex4-1}

\usepackage{graphicx}
\usepackage{dcolumn}
\usepackage{bm}
\usepackage{amsmath}
\usepackage{xcolor}
\usepackage{colortbl}
\usepackage{pifont}
\usepackage{verbatim}
\usepackage{natbib}
\setcitestyle{square,numbers}
\usepackage[caption=false]{subfig}
\usepackage{mathtools}
\usepackage{comment}
\usepackage[utf8]{inputenc}
\usepackage{hyperref}

\begin{document}

\title{Prospects and limitations for constraining light relics with primordial abundance measurements}

\author{Alex Lagu\"e}
\affiliation{Department of Astronomy \& Astrophysics, University of Toronto, 50 St. George St., Toronto, ON M5S 3H4, Canada,
\\
Canadian Institute for Theoretical Astrophysics, University of Toronto, 60 St. George Street, 14th floor
Toronto, ON. M5S 3H8, Canada,
\\
Dunlap Institute for Astronomy \& Astrophysics, University of Toronto, 50 St. George Street, Toronto, ON M5S 3H4, Canada.}
\email{lague@astro.utoronto.ca}
\homepage{https://www.cita.utoronto.ca/~lague/}
\author{Joel Meyers}%
\affiliation{Southern Methodist University, 3215 Daniel Ave., Dallas, TX 75275, USA}%

\date{\today}

\begin{abstract}
The light relic density affects the thermal and expansion history of the early Universe leaving a number of observable imprints.  We focus on the primordial abundances of light elements produced during the process of Big Bang nucleosynthesis which are influenced by the light relic density.  Primordial abundances can be used to infer the density of light relics and thereby serve as a probe of physics beyond the Standard Model.  We calculate the observational uncertainty on primordial light element abundances and associated quantities that would be required in order for these measurements to achieve sensitivity to the light relic density comparable to that anticipated from upcoming cosmic microwave background surveys.  We identify the nuclear reaction rates that need to be better measured to maximize the utility of future observations. We show that improved measurements of the primordial helium-4 abundance can improve constraints on light relics, while more precise measurements of the primordial deuterium abundance are unlikely to be competitive with cosmic microwave background measurements of the light relic density. 

\end{abstract}

\maketitle

\section{Introduction}\label{Sec:Intro}
In the first few minutes of the evolution of the Universe, a handful of light elements were synthesized in a process known as Big Bang nucleosynthesis (BBN); see~\cite{eta_measure} for a recent review. The resulting abundances of these elements set the stage for the subsequent synthesis of heavier elements in stars and supernovae~\cite{Nomoto:2013oal}.  BBN is one of the cornerstones of modern cosmology.  The successes of the theory, along with measurements of the cosmic microwave background (CMB) and observations of the distances and redshifts of distant galaxies, have firmly established the hot Big Bang model of the Universe.

BBN is a process which depends on all four fundamental forces and has only a single free parameter, the baryon-to-photon ratio $\eta$~\cite{eta_measure}, which itself can be measured by other means including galaxy surveys~\cite{Percival:2001hw} and CMB measurements~\cite{2018planck}.  
As such, BBN has long provided a useful constraint on models of physics beyond the Standard Model~\cite{Iocco:2008va}, and it will continue to be a powerful tool as data and computational power improve~\cite{Grohs:2019cae}.

The BBN process comprises a chain of nuclear reactions and is highly dependent on the composition and expansion rate of the early Universe.
Since three-body and four-body interactions were exceedingly rare during BBN, only the lightest few nuclei were formed in the process.
The relatively low binding energy of deuterium~\cite{H2unstable} and the very high density of photons relative to baryons delayed the formation of tritium, helium, and heavier nuclei until the temperature dropped well below their binding energies and some fraction of free neutrons had decayed; this delay is known as the deuterium bottleneck.
The baryon-to-photon ratio, $\eta$, plays an important role in determining when the deuterium bottleneck is overcome, making way for the rest of the BBN processes. If $\eta$ is high, deuterium was formed earlier due to the lower density of high energy photons capable of dissociating deuterium, leaving less time for neutron decay, thereby leading to a higher helium yield and lower unburned deuterium abundance. Conversely, if $\eta$ is low, the higher density of photons
delayed the synthesis of helium and heavier elements. This sensitivity to $\eta$ makes deuterium a very good baryometer~\cite{eta_measure}.

The Hubble expansion rate, $H$, is determined by the Friedmann equation, $3H^2 = 8\pi G \rho$, where $\rho$ is the the energy density of the Universe which is dominated by the radiation density during BBN, $\rho= \rho_r$. It is conventional to describe the total radiation density after neutrino decoupling in terms of the photon temperature as \cite{neutrinoPLANCK,cmbs4}
\begin{align}
\rho_r = \frac{\pi^2 k_B^4}{15 \hbar^3 c^3}\bigg[1+ \frac{7}{8}\bigg(\frac{4}{11}\bigg)^{4/3} N_{\mathrm{eff}} \bigg] T^4_{\gamma} \, ,
\end{align}
where $T_\gamma$ is the photon temperature and $N_{\mathrm{eff}}$ is the effective number of neutrino species which measures the total density of light relics. The expansion rate impacts the relationship between temperature and time in the early Universe, affecting the neutron-to-proton ratio at weak freeze-out, the time for free neutron decay, and ultimately, the primordial abundances~\cite{eta_measure}.

$N_\mathrm{eff}$ is an especially interesting parameter since it acts as a direct window into fundamental physics~\cite{Green:2019glg}.
Modifications to the list of ingredients of our Universe, the interactions between them, or to the thermal history of the early Universe can all manifest as changes to $N_\mathrm{eff}$. 
New light states are among the ingredients in many extensions of the Standard Model.  The extreme conditions present in the early Universe were sufficient to bring even very weakly interacting particles into thermal equilibrium.  Long-lived light particles produced in the early universe contribute to the radiation density and alter $N_\mathrm{eff}$.  In the Standard Model of particle physics and assuming a standard thermal history of the early universe, $N_\mathrm{eff}$ just measures the energy density in the cosmic neutrino background and is given by $N_{\mathrm{eff}}^\mathrm{SM}=3.046$~\cite{Mangano:2005cc,Grohs:2015tfy,deSalas:2016ztq}. Note that even in the standard case, $N_\mathrm{eff}$ is not an integer due to the fact that neutrino decoupling is not instantaneous and the cosmic neutrino background does not have a perfectly thermal distribution. It is convenient to define the deviation of the light relic density from the Standard Model value as $\Delta N_{\mathrm{eff}} \equiv  N_{\mathrm{eff}} - N_{\mathrm{eff}}^{\mathrm{SM}}$. 
A measured value greater than the Standard Model prediction, $\Delta N_{\mathrm{eff}} >0$, could indicate the presence of new light relics such as light sterile neutrinos, a thermal population of axions, or some other form of dark radiation~\cite{Essig:2013lka,Marsh:2015xka,Alexander:2016aln,Abazajian:2017tcc,Green:2019glg}.

For light thermal relics, the contribution to $\Delta N_\mathrm{eff}$ can be computed from the spin and decoupling temperature of the new species~\cite{Green:2019glg}
\begin{align} 
    \Delta N_\mathrm{eff} = g_s \left(\frac{43/4}{g_\star(T_\mathrm{F})}\right)^{4/3} \, ,
    \label{eq:Delta_Neff}
\end{align}
where $g_s$ is the effective number of spin states of the new species including a factor $7/8$ for fermions, $g_\star$ is the effective number of relativistic degrees of freedom contributing to expansion, and $T_\mathrm{F}$ is the temperature at which the new species was last in equilibrium with the thermal plasma.
Interestingly, models which contain just a single new species beyond the contents of the Standard Model predict clear thresholds for $\Delta N_\mathrm{eff}$.  For example, a value of $\Delta N_{\mathrm{eff}} \geq 0.047$ is predicted by models containing a thermal population of light fermions, while a smaller bound of $\Delta N_{\mathrm{eff}} \geq 0.027$ governs any model in which light scalars such as axions or axion-like particles were in equilibrium at any point in the early Universe~\cite{cmbs4,Green:2019glg}.

The next generation of cosmological observations is entering a regime where even very small deviations from the Standard Model prediction for $N_\mathrm{eff}$ will be tightly constrained~\cite{Green:2019glg}. 
The improvements from CMB observations will be especially impressive and are anticipated over the next several years with planned and proposed experiments like Simons Observatory~\cite{Ade:2018sbj,Abitbol:2019nhf}, CMB-S4~\cite{cmbs4,Abazajian:2019eic}, PICO~\cite{Hanany:2019lle}, and CMB-HD~\cite{Sehgal:2019ewc}.

Here we investigate what would be required of observations of primordial light element abundances (and associated quantities) to achieve a sensitivity to the light relic density $N_\mathrm{eff}$ comparable to that anticipated from upcoming CMB observations.  This study is motivated in several ways.  First, current limits on $N_\mathrm{eff}$ inferred from the CMB~\cite{2018planck} and from light element abundances~\cite{eta_measure,Yp_ab,Cooke:2017cwo} have a comparable uncertainty and agree well.  These two observational schemes for measuring $N_\mathrm{eff}$ are subject to different systematic uncertainties and potential biases and thereby provide a useful cross check on one another.  We will show how much abundance measurements need to improve in order to maintain this utility as CMB observations steadily improve $N_\mathrm{eff}$ constraints in the coming years.  Next, light element abundances and CMB observations are sensitive to slightly different aspects of the light relic density.  For example, BBN is influenced differently by changes to the density of cosmic neutrinos than by the addition of some dark radiation~\cite{eta_measure}.  The CMB power spectrum is impacted by both the mean density of light relics and by fluctuations in the light relic density~\cite{Bashinsky:2003tk,Baumann:2015rya}.  Furthermore, BBN is impacted by the light relic density at times earlier than those which influence the CMB power spectrum.  Differences in the value of $N_\mathrm{eff}$ inferred from primordial light elements and that from the CMB may indicate a modified thermal history or other new physics~\cite{Fischler:2010xz,Cadamuro:2010cz,Menestrina:2011mz,Hooper:2011aj,Millea:2015qra}.

We choose to focus here on the light relic density since it is a quantity with clear theoretical thresholds and which can be directly compared with upcoming constraints from the CMB.  However, there is a great deal that can be inferred from measurements of primordial abundances that need not have anything to do with the light relic density; see~\cite{Iocco:2008va,Fields:2011zzb,eta_measure,Adams:2017dii,dePutter:2018xte,Grohs:2019cae} for some examples.  Even for these other applications, some of our results are still applicable.  The degeneracies of $N_\mathrm{eff}$ with nuclear rates and with the baryon-to-photon ratio that we discuss below can limit the ability to infer any parameter which influences BBN.  Therefore, while we will frame our discussion in terms of $N_\mathrm{eff}$, some of the lessons will apply more broadly to BBN science.  Specifically, as the uncertainty on the primary abundances improves, the limitations imposed by the uncertainty on nuclear rates and on the baryon-to-photon ratio will need to be addressed to fully utilize forthcoming observational data.


\begin{table}[t]
\centering
\begin{tabular}{c c c}
  \hline
  \hline
  Observable & Current Constraint & Reference \\ \hline
  $\eta$ & $(6.12 \pm 0.04)\times 10^{-10}$  & \cite{2018planck} \\ 
  $Y_\mathrm{p}$ & $0.2449 \pm 0.0040$  & \cite{Yp_ab} \\
  D/H & $(2.527\pm 0.03)\times 10^{-5}$ & \cite{Cooke:2017cwo} \\
  \textsuperscript{3}He/H & $(1.1\pm 0.2)\times 10^{-5}$  & \cite{He3_ab} \\
  \textsuperscript{7}Li/H & $(1.23 \pm 0.68) \times 10^{-10}$  & \cite{Li_ab}\\
  \textsuperscript{6}Li/H & $\sim 10^{-5}$ \textsuperscript{7}Li/H  & \cite{lithium6} \\
  \hline
  \hline
\end{tabular}
\caption{Current observational constraints on the baryon-to-photon ratio and the primordial abundances.
\label{tab:obs}}
\end{table}



\section{Forecasts} \label{Sec:Proposal} 
In order to forecast the constraining power of future measurements of primordial light element abundances, we compute the Fisher matrix,
\begin{align} \label{eq:Fisher}
    F_{ij} = \sum_a \frac{\partial X_a}{\partial p_i} \sigma^{-2}_a \frac{\partial X_a}{\partial p_j} \, ,
\end{align}
where the $X_a$ are the primordial abundances of of \textsuperscript{2}H, \textsuperscript{3}He, \textsuperscript{4}He (or equivalently $Y_\mathrm{p}\equiv\rho(^{4}\mathrm{He})/\rho_b$), \textsuperscript{6}Li, and \textsuperscript{7}Li whose measurement errors $\sigma_a$ we take to be independent.  The set of parameters $p_i$ is composed of the effective number of neutrino species $N_{\mathrm{eff}}$ (which we wish to constrain) along with a set of secondary parameters, namely the baryon-to-photon ratio ($\eta\equiv n_b/n_\gamma$) and the rates of one hundred nuclear reactions involved in the BBN process (the set of nuclear rates we consider is shown in Table~\ref{tab:nuc_rates}).  
This choice of parameters will allow us to identify the degeneracies with $N_\mathrm{eff}$ that will limit our ability to infer the light relic density from future measurements of primordial abundances.

For a given set of input parameters, we compute the abundances using a modified version of the code \texttt{AlterBBN}~\cite{alterBBN,Arbey:2018zfh}. \texttt{AlterBBN} is a publicly available\footnote{\url{https://alterbbn.hepforge.org}} C program which rapidly computes the abundances of primordial elements given a set of cosmological parameters including $\eta$ and $N_\mathrm{eff}$. The modification which we have introduced is to allow for changes to the rates of nuclear reactions so that the user can scale any given nuclear rate, thus making them adjustable input parameters. 
Derivatives of the abundances with respect to each parameter are computed using finite difference with step sizes chosen to ensure numerical stability.

The current observational constraints on the primordial abundances are summarized in Table \ref{tab:obs}.  In the next section we will show how improvements from future measurements of these abundances will constrain the light relic density, $N_\mathrm{eff}$.  As we will see, such improvements would be limited by the current uncertainties in $\eta$ and the rates of some nuclear reactions relevant for BBN, so we consider the possibility of improved measurements of these quantities as well.

In what follows we will show results for various priors on the baryon-to-photon ratio $\eta$. The current bound derived from \textit{Planck} CMB data is 
$\eta = (6.12\pm 0.04)\times 10^{-10}$~\cite{2018planck}.
The uncertainty on $\eta$ anticipated from an experiment like CMB-S4 is $\sigma(\eta)/\eta = 0.0023$~\cite{cmbs4,Abazajian:2019eic} (for a $\Lambda$CDM$+N_\mathrm{eff}+\sum m_\nu$ cosmology).\footnote{We will refer throughout to uncertainties expected from CMB-S4, assuming a configuration consistent with the reference design shown in~\cite{Abazajian:2019eic}, though the forecasted errors for the relevant parameters from an experiment like PICO are similar.}  We also consider the case of no $\eta$ prior, a case that requires the measurement of at least two primordial abundances to infer the light relic density.

The current uncertainties on the nuclear rates we consider are tabulated in Table~\ref{tab:nuc_rates} of the appendix with a numbering which follows the nuclear reaction order of the latest version of the \texttt{AlterBBN} code~\cite{alterBBN,Arbey:2018zfh}. 
Many of the uncertainties on the nuclear rates were obtained from the NACRE II compilation of astrophysical nuclear reactions \cite{NACREII}. The accepted value and the uncertainty on any given rate depend on the temperature.  We normalize the rates and uncertainties at the fixed temperature of $T=10^{10}$~K.  We take the fractional error on each rate to be given by half the interval between the high and low rates divided by central value. For rates whose uncertainties are not tabulated elsewhere, we conservatively assign a $15\%$ uncertainty.  None of the unlisted rates are currently the limiting factor for the prediction of any abundance we consider here.  
Reduced uncertainty on some of these rates can be anticipated in the coming years from experiments like LUNA~\cite{luna}.


\section{Results}
\label{sec:results}

\begin{figure}[t]
    \centering
    \includegraphics[width=.9\linewidth]{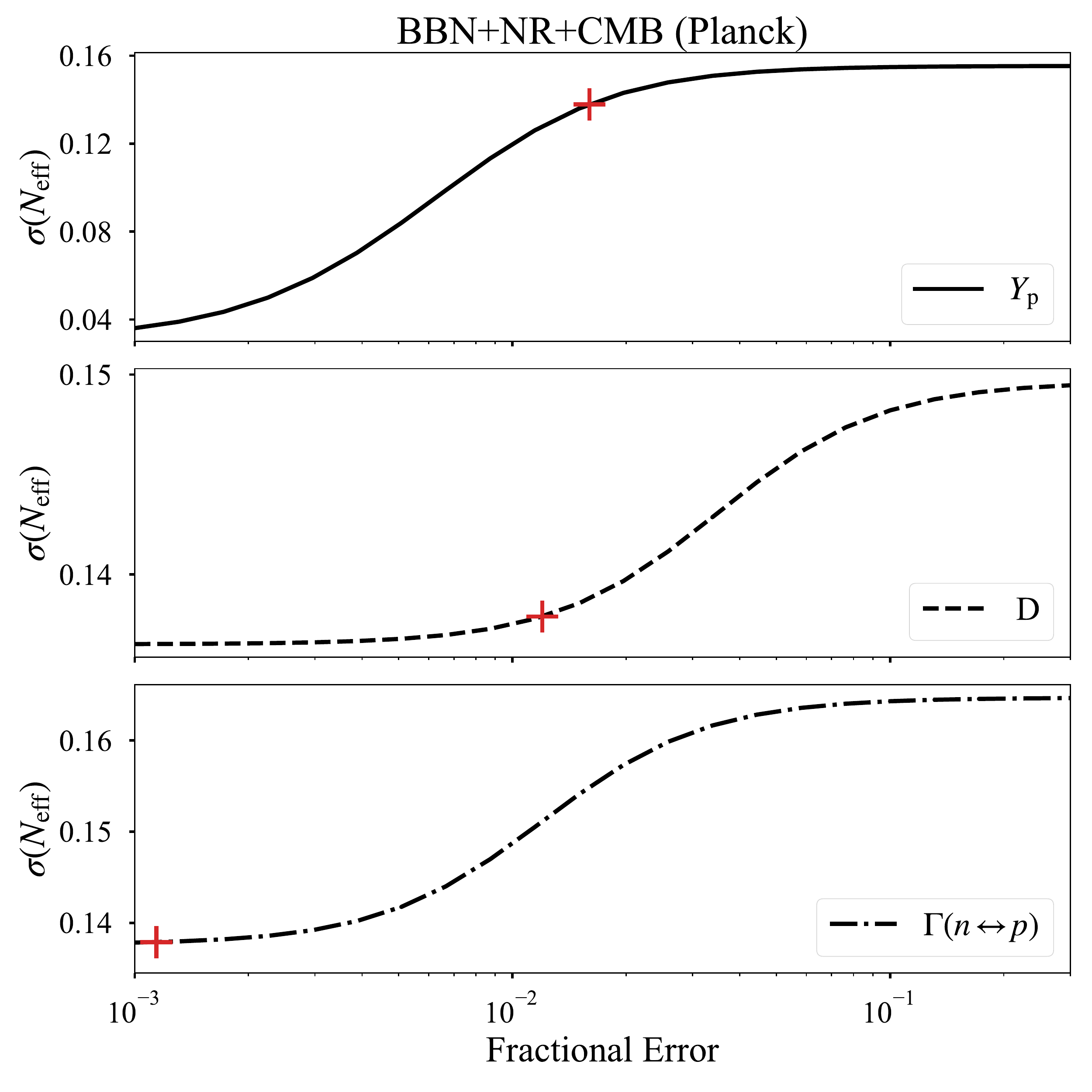}
    \caption{Forecasted 1-$\sigma$ error on $N_\mathrm{eff}$ as a function of the observational uncertainty of the primordial deuterium abundance, the primordial \textsuperscript{4}He abundance ($Y_\mathrm{p}$), and the neutron lifetime using current constraints on $\eta$ and $N_\mathrm{eff}$ from \textit{Planck}~\cite{2018planck}, and current primordial abundance measurements listed in Table~\ref{tab:obs}, marginalized over all nuclear rates with uncertainties listed in Table~\ref{tab:nuc_rates}. The red crosses indicate the current uncertainties.}
    \label{fig:Planck}
\end{figure}

We now explore how improved observations of primordial abundances would affect constraints on the light relic density, both alone and in combination with current and future CMB measurements.
We begin by analyzing the effect of more stringent bounds on the abundances of deuterium and \textsuperscript{4}He when combined with existing CMB data from \textit{Planck}. In Figure~\ref{fig:Planck}, we show the resulting value of $\sigma(N_{\mathrm{eff}})$ as a function of the fractional uncertainty on the helium abundance, deuterium abundance, and neutron lifetime assuming the current observational bounds on all other parameters, taking in particular the current \textit{Planck} bounds on $\eta$ and $N_\mathrm{eff}$.  While an improved measurement of $Y_\mathrm{p}$ could provide significantly tighter constraints on $N_\mathrm{eff}$, we see that a tighter deuterium constraint leads to little improvement for $N_\mathrm{eff}$.  We will explore the reasons for this behavior below.  We also find that better measurements of the neutron lifetime would not improve constraints on $N_\mathrm{eff}$ and such measurements would limit constraining power only if the uncertainty were about an order of magnitude larger than those of the current measurement.

\begin{figure}[t]
    \centering
    \includegraphics[width=.9\linewidth]{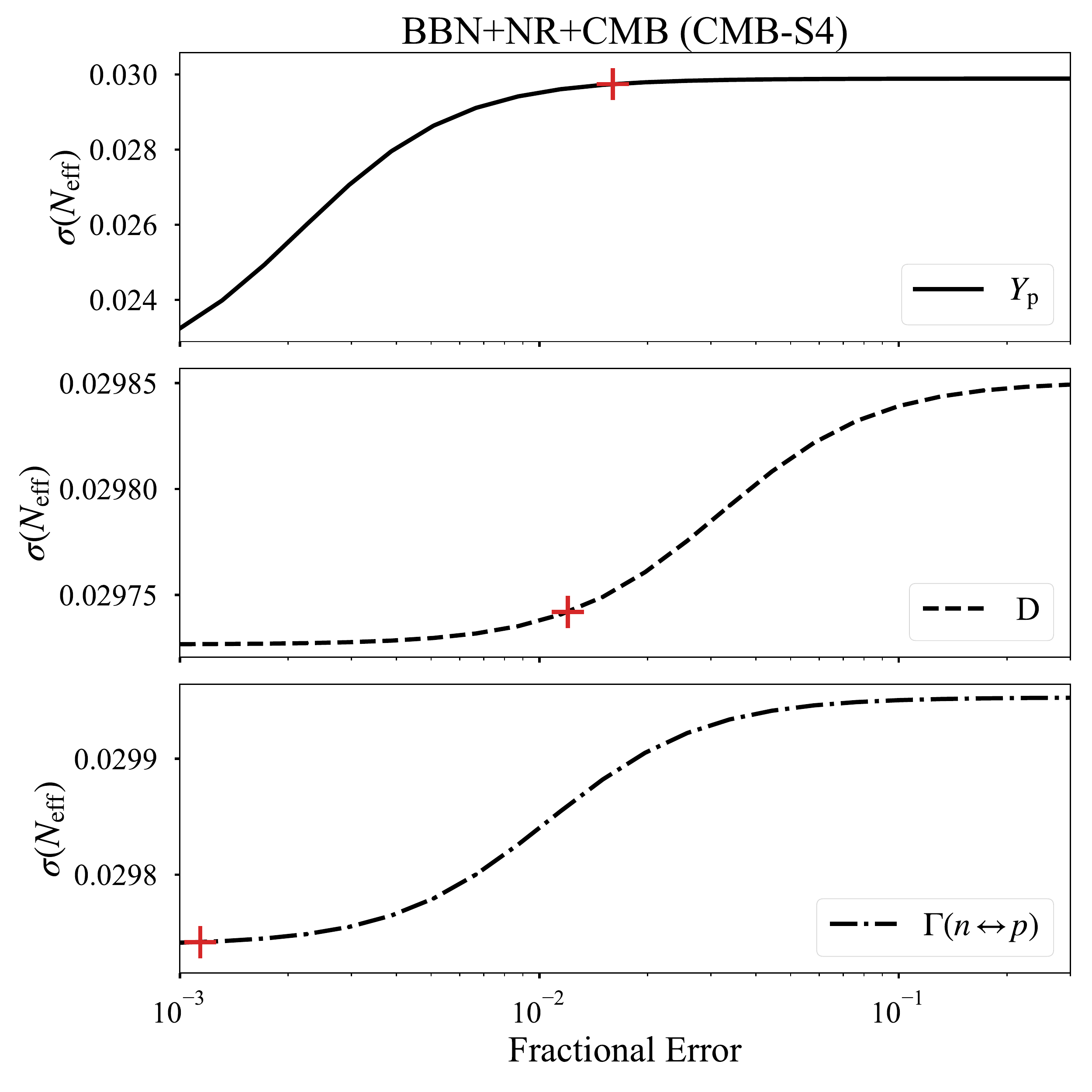}
    \caption{Same as Figure~\ref{fig:Planck} but with forecasted constraints on $\eta$ and $N_\mathrm{eff}$ from CMB-S4~\cite{cmbs4,Abazajian:2019eic}. }
    \label{fig:S4}
\end{figure}

A similar pattern is seen in Figure~\ref{fig:S4} which shows the same results as in Figure~\ref{fig:Planck} but using forecasted errors for $N_\mathrm{eff}$ and $\eta$ from CMB-S4 rather than current measurements from \textit{Planck}. 
In this case, improved measurements of $Y_\mathrm{p}$ could still improve constraints on $N_\mathrm{eff}$, but the improvement is less significant since the error is dominated by the CMB-only measurement of $\sigma(N_\mathrm{eff})=0.03$ that CMB-S4 will provide~\cite{cmbs4,Abazajian:2019eic}.\footnote{When quoting constraints on $N_\mathrm{eff}$ from CMB experiments here and elsewhere, we are assuming that $Y_\mathrm{p}$ is fixed to be consistent with predictions of BBN (as a function of $\eta$ and $N_\mathrm{eff}$).  Modifications to the process of BBN or other new physics could alter the relationship between $Y_\mathrm{p}$ and $N_\mathrm{eff}$, but such scenarios are outside the scope of this work.}  Improved deuterium measurements would have little impact in this case, and the forecasts are mostly insensitive to the uncertainty in the neutron lifetime.

From Figure~\ref{fig:Planck}, we observe that with the current uncertainties on $\eta$ and on the relevant nuclear rates, improvements in the measurement of the primordial \textsuperscript{4}He abundance, $Y_\mathrm{p}$, provide the largest benefit for constraints on $N_{\mathrm{eff}}$. Measurements of $Y_\mathrm{p}$ which improve on current bounds by an order of magnitude could yield constraints comparable to (but not quite reaching) the bound projected by CMB-S4 of $\sigma(N_\mathrm{eff})=0.03$. 
Improved measurements of the primordial deuterium abundance fail to yield significant improvements in the constraint on $N_\mathrm{eff}$.  This is due primarily to a degeneracy with the rate of the reaction \textsuperscript{2}H($p$, $\gamma$)\textsuperscript{3}He, listed as nuclear rate 20 in Table~\ref{tab:nuc_rates}~\cite{eta_measure,2018planck}.  Below we will explore how improved measurements of this and other nuclear rates affect the constraining power of abundance measurements.  We do not show results for constraints inferred from \textsuperscript{3}He or lithium, since they provide a negligible benefit for $N_\mathrm{eff}$. 

The accepted measurement of the neutron lifetime has changed quite dramatically over the years~\cite{cmbs4}. The value of the lifetime chosen for this study (which is in agreement with the value used in \texttt{AlterBBN}) is that of $\tau_n=880.2 \pm 1.0$~\cite{neutrondecay} which translates to a fractional error on the $n\leftrightarrow p$ nuclear rate of 0.00114. The impact on $\sigma(N_{\mathrm{eff}})$ of varying the uncertainty on $\tau_n$  is displayed in the third rows of Figures \ref{fig:Planck} and \ref{fig:S4}. In both instances, we see that improvements on the already very small uncertainty would not improve the constraining power for $N_\mathrm{eff}$.  We will show below that the uncertainty on the neutron lifetime could become the limiting factor for the inference of $N_\mathrm{eff}$ if the measurements of $Y_\mathrm{p}$ are improved by more than an order of magnitude.


\begin{table}[t]
\centering
\begin{tabular}{ c c c c c c}
  \hline
  \hline
  Rank & Prior $\eta$ & Prior $N_{\mathrm{eff}}$ & $\sigma(Y_\mathrm{p})$ & $\sigma(\mathrm{D/H})$ & $\sigma(N_{\mathrm{eff}})$\\ \hline
  1 & CMB-S4          & CMB-S4          & 0.16\% & 1.2\% & 0.0246   \\
  2 & CMB-S4          & CMB-S4          & 1.6\% & 1.2\% & 0.0296  \\
  3 & \textit{Planck} & \textit{Planck} & 0.16\% & 1.2\% & 0.0422  \\
  4 & CMB-S4 & None & 0.16\% & 0.12\% & 0.0433  \\
  5 & \textit{Planck} & None & 0.16\% & 0.12\%  & 0.0434  \\
  6 & \textit{Planck} & None & 0.16\% & 1.2\%  & 0.0435  \\
  7 & \textit{Planck} & \textit{Planck} & 1.6\% & 1.2\% & 0.138   \\
  8 & \textit{Planck} & None & 1.6\% & 0.12\%  & 0.222  \\
  9 & CMB-S4 & None & 1.6\% & 1.2\%  & 0.226  \\
  10 & \textit{Planck} & None & 1.6\% & 1.2\%  & 0.228  \\
  
  \hline
  \hline
\end{tabular}
\caption{Rankings of approaches to reduce the uncertainty on $N_{\mathrm{eff}}$ for various future CMB and abundance measurements, considering possible order-of-magnitude improvements on the abundance measurements of helium and deuterium. \label{tab:Yp_CMB-S4}}
\end{table}


In Table~\ref{tab:Yp_CMB-S4}, we summarize the results of Figures~\ref{fig:Planck} and \ref{fig:S4} by considering specific bounds on $\sigma(Y_\mathrm{p})$ and
$\sigma(\mathrm{D/H})$ with \textit{Planck} and CMB-S4 priors on $\eta$ and $\sigma(N_{\mathrm{eff}})$ and assuming current uncertainties on all nuclear rates. Ten such cases are considered and ranked based on the constraints on $N_{\mathrm{eff}}$.  The largest improvements come from using the forecasted CMB-S4 error on $N_\mathrm{eff}$.  Improving measurements $Y_\mathrm{p}$ by an order of magnitude provides constraints on $N_\mathrm{eff}$ that are comparable with that achieved by CMB-S4, though they are in each case larger by more than $40\%$.  There are insignificant improvements from better deuterium measurements when including CMB-S4, so these cases are not listed in Table \ref{tab:Yp_CMB-S4}. Finally, the most optimistic scenario we explore for the constraint on the light relic density occurs when combining the CMB-S4 priors with an order-of-magnitude improvement on the primordial helium measurement where a constraint of $\sigma(N_\mathrm{eff}) = 0.0246$ is attained. With such a measurement, one would be able to identify the presence of any additional light thermal fermions (which predict $\Delta N_{\mathrm{eff}}\geq 0.047$) at just below 2-$\sigma$ confidence.

We now turn to identifying why the constraints on the light relic density saturate for significantly improved primordial abundance measurements, a task that is necessary to maximize the utility of future observations. Considering each abundance in isolation, we calculate the correlation coefficients of $N_\mathrm{eff}$ with each of $\eta$ and the 100 nuclear rates considered in our BBN network.  This allows us to identify which parameters are most degenerate with $N_\mathrm{eff}$ for each abundance.


\begin{figure*}[t!]
    \centering
    \includegraphics[width=1.\linewidth]{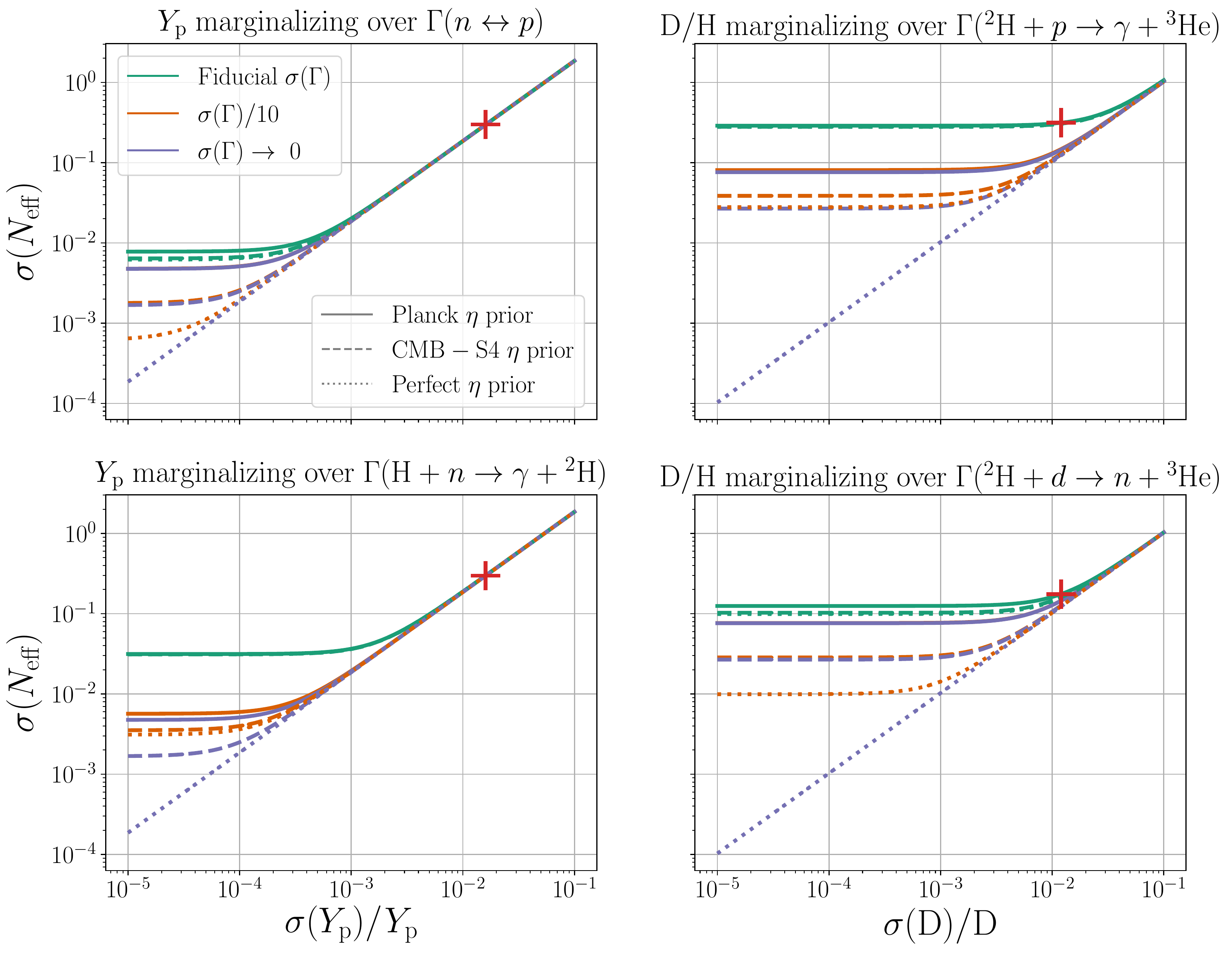}
    \caption{Forecasted 1-$\sigma$ errors for $N_\mathrm{eff}$ as a function of  $\sigma(Y_\mathrm{p})/Y_\mathrm{p}$ and $\sigma(\mathrm{D})/\mathrm{D}$ marginalizing over $\eta$ and a specified nuclear rate (with all other parameters held fixed). The red cross denotes the current uncertainties. }
    \label{fig:NR_eta}
\end{figure*}


We find the baryon-to-photon ratio $\eta$ to be quite strongly correlated with both the deuterium and helium abundances. For each abundance the most highly correlated nuclear rates are typically those directly leading to their formation or those involved in their decay. A positive correlation with $N_\mathrm{eff}$ means that the effect of increasing the value of that nuclear rate is degenerate with increasing the value of $N_\mathrm{eff}$, even if that corresponds physically to a decrease in value of the primordial abundance. An example of such a correlation is that between the rate of \textsuperscript{2}H($p$,$\gamma$)\textsuperscript{3}He and $N_\mathrm{eff}$ for the production of deuterium. Increasing $N_\mathrm{eff}$ leads to an increase in the radiation density in the Universe which results in an increase in the expansion rate and to a faster transition out of the deuterium bottleneck. The deuterium can then be more efficiently converted into other elements, a consequence which is also reached if one simply increases the rate of proton capture by deuterium. 

In Figure~\ref{fig:NR_eta} we show how the various degeneracies of $N_\mathrm{eff}$ with the most relevant nuclear rates and with $\eta$ impact the constraining power of future measurements of \textsuperscript{4}He and deuterium.  In the case of \textsuperscript{4}He, the neutron decay rate is by far the most strongly correlated with $N_\mathrm{eff}$, but the higher uncertainty on the rate of H($n$,$\gamma$)\textsuperscript{2}H implies that the latter is a more significant degeneracy.  Current constraints on the baryon-to-photon ratio will not limit the inference of $N_\mathrm{eff}$ from $Y_\mathrm{p}$ unless primordial helium abundance measurements improve my more than an order of magnitude. 

For deuterium, we see that the significant correlation between the rate of \textsuperscript{2}H$(p,\gamma)$\textsuperscript{3}He (nuclear rate 20 in Table~\ref{tab:nuc_rates}) and $N_\mathrm{eff}$ leads to a degeneracy which limits the constraining power of even current measurements of the deuterium abundance~\cite{eta_measure,Cooke:2017cwo,2018planck}. The rate of \textsuperscript{2}H$(d,n)$\textsuperscript{3}He (nuclear rate 28) is also fairly degenerate with $N_\mathrm{eff}$ in its effects on the deuterium abundance. In order to realize the full potential of current and improved measurements of the primordial deuterium abundance, it will be necessary to improve measurements of the rates of these reactions as envisioned by the upcoming updated LUNA experiment~\cite{luna}.

Even if we fix all relevant nuclear rates, the degenerate impacts of $\eta$ and $N_\mathrm{eff}$ on the primordial deuterium abundance are beginning to become important with the current observational uncertainties.  Therefore, in order for future deuterium abundance measurements to provide significant improvements on the inference of the light relic density, both better nuclear rate measurements and better measurements of the baryon-to-photon ratio are required.  Since the latter is most likely to come from upcoming observations of the CMB, and those same measurements will provide very tight constraints on $N_\mathrm{eff}$, we can conclude that primordial deuterium measurements in isolation are unlikely to be competitive with future CMB measurements for constraining the light relic density.  As discussed above, however, there are myriad applications for measurements of primordial abundances that do not directly involve the light relic density, and which will greatly benefit from improved measurements of deuterium.


\begin{figure*}[t!]
    \begin{center}
    \includegraphics[width=\textwidth]{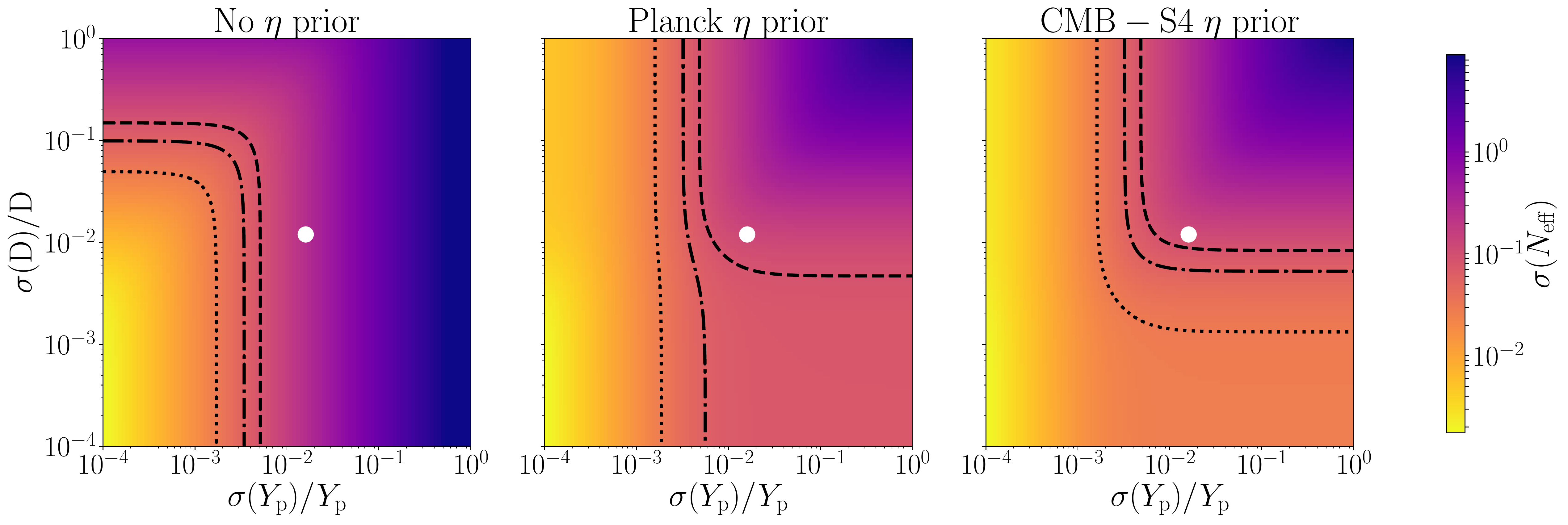}
    \caption{Forecasted 1-$\sigma$ errors on $N_\mathrm{eff}$ from joint constraints on primordial deuterium and \textsuperscript{4}He abundances with all nuclear reaction rates held fixed for three choices of CMB priors on $\eta$: no CMB prior (\textit{left}), the current uncertainty from \textit{Planck} measurements (\textit{center}), and the projected precision from CMB-S4 (\textit{right}). The dotted, dot-dashed and dashed lines represent the $\sigma(N_\mathrm{eff}) = 0.03,\; 0.06,$ and $0.09$ contours respectively while the white dots denote the current uncertainties on the primordial abundances. 
    }\label{fig:color_plots}
    \end{center}
    \vspace{.1cm}
\end{figure*}


We also consider the possibility of using only abundance measurements to constraint the light relic density, without relying on improved CMB measurements of the baryon-to-photon ratio.  In this case, measurements of at least two abundances are required, since for any single abundance, there is a perfect degeneracy between $N_\mathrm{eff}$ and $\eta$.  In Figure~\ref{fig:color_plots}, we show how joint constraints on the primordial abundances of \textsuperscript{4}He and deuterium translate into constraints on $N_\mathrm{eff}$ with all nuclear rates fixed.

Figure~\ref{fig:color_plots} shows that, given the current uncertainty in the deuterium abundance, improved measurements of the primordial \textsuperscript{4}He abundance can lead to significant improvements in the inference of $N_\mathrm{eff}$, independent of any CMB measurements.  The message is less straightforward for deuterium. In the left panel of Figure~\ref{fig:color_plots}, we see that in the absence of a prior on $\eta$, only $Y_\mathrm{p}$ can be used to improve constraints on $N_\mathrm{eff}$, while improved deuterium measurements give no additional information on the light relic density. In the center panel, we see that when including the current measurement of $\eta$ from \textit{Planck} data, there is room for minor improvement in $N_\mathrm{eff}$ from better deuterium measurements (assuming the uncertainties in the relevant unclear rates can be reduced).  The right panel shows that with the measurement of $\eta$ anticipated from CMB-S4, improved measurements of the primordial deuterium abundance can lead to a more precise measurement of $N_\mathrm{eff}$, ultimately giving a constraint about a factor of two better than that expected from CMB-S4 (see also Figure~\ref{fig:NR_eta}).


\section{Discussion}

The light relic density is a particularly useful cosmological observable due to the broad implications its measurement has for physics beyond the Standard Model.  Current measurements of the light relic density from primordial abundances and from the CMB have comparable uncertainties.  Future measurements of the CMB (and also of large-scale structure~\cite{Baumann:2017gkg}) will greatly improve the precision with which we measure the density of light relics.  Here we showed what would be required of primordial abundance measurements to keep pace with the rapidly improving CMB constraints.

We found that roughly order-of-magnitude improvements in the measurement of the primordial abundance of \textsuperscript{4}He or deuterium would be necessary to infer $N_\mathrm{eff}$ with an uncertainty comparable to that expected from CMB-S4 or PICO, $\sigma(N_\mathrm{eff})=0.03$.  In the case of deuterium, it would be additionally required that the uncertainties in the rates of the nuclear reactions \textsuperscript{2}H$(p,\gamma)$\textsuperscript{3}He and \textsuperscript{2}H$(d,n)$\textsuperscript{3}He be improved by about an order of magnitude, and that the uncertainty on the baryon-to-photon ratio be reduced to a level expected from the measurement by a CMB survey like CMB-S4.

The limitations on the improvements expected from more precise deuterium measurements are somewhat disappointing, since the precision with which the primordial deuterium abundance has been measured has greatly improved in recent years~\cite{eta_measure,deuterium_ab,Cooke:2016rky,Cooke:2017cwo}.  These measurements are primarily made through observation of quasar absorption systems, and the constraint is dominated by just a few of the most pristine systems. Further improvements in deuterium measurements should be possible in the coming years, especially as 30m class optical and near infrared telescopes are constructed, which should be able to detect and analyze many more quasars~\cite{Maiolino:2013bsa,Cooke:2016rky,Grohs:2019cae}.

The current best measurements of the primordial \textsuperscript{4}He abundance come from measurements of nearby metal-poor dwarf galaxies~\cite{Izotov:2014fga,Yp_ab}.  These measurements are limited by systematic uncertainties rather than statistics, making it difficult to determine how errors might improve in the future.  An alternative method to measure the primordial helium abundance using quasar absorption spectra has recently been developed but is not yet competitive with galactic measurements~\cite{Cooke:2018qzw}.

The CMB is sensitive to the primordial helium density due to its impact on the number density of free electrons around recombination, which affects the damping tail of the CMB power spectrum~\cite{Hu:1996mn}.  Changes to the light relic density also impact the CMB damping tail but additionally produce other effects which allows the parameters to be simultaneously constrained~\cite{Bashinsky:2003tk,Hou:2011ec,Baumann:2015rya}.  The constraints on $Y_\mathrm{p}$ that are expected from CMB-S4 when both $N_\mathrm{eff}$ and $Y_\mathrm{p}$ are free to vary are comparable to current astrophysical uncertainties~\cite{cmbs4}.  On the other hand, standard BBN predicts a specific relationship between $N_\mathrm{eff}$ and $Y_\mathrm{p}$, and CMB constraints on $N_\mathrm{eff}$ are tighter when BBN consistency is imposed~\cite{2018planck,cmbs4}.  Therefore, if the goal is to obtain the tightest constraint on the light relic density (assuming it is constant during the relevant epochs), direct CMB constraints on $N_\mathrm{eff}$ will always be a better strategy than using a CMB inference of $Y_\mathrm{p}$.

While our results show in part that CMB measurements are likely to be the most promising path forward to improve the measurements of the light relic density, there is a wide range of applications for primordial abundance measurements both alone and in combination with CMB measurements which ensures that both sets of measurements will be extremely valuable to the future of cosmology~\cite{Iocco:2008va,eta_measure,dePutter:2018xte,Green:2019glg,Grohs:2019cae}.


\section*{Acknowledgments}
We are grateful to Ren\'{e}e Hlo\v{z}ek, Chris Matzner, and Alexander van Engelen for helpful conversations.
JM is supported by the US Department of Energy under grant no.~DE-SC0010129.


\section*{Appendix: Table of Nuclear Rates}
Below, we present the table of nuclear rates along the associated uncertainty used in this work.
\begin{table}[h]
\centering
\begin{tabular}{c c c c | c c c c}
  \hline
  \hline
  Rate & Reaction & Error (\%) & Reference & Rate & Reaction & Error (\%) & Reference \\ \hline
   1 & $n \leftrightarrow p$ & 0.114 & \cite{neutrondecay} &
   2 & \textsuperscript{3}H $\to e^- + \nu+$ \textsuperscript{3}He & 0.393 & \cite{tritium_decay} \\
   3 & \textsuperscript{8}Li $\to e^- + \nu+$ 2\textsuperscript{4}He & 0.726 & \cite{weak5-10} &
   4 & \textsuperscript{12}B $\to e^- + \nu+$ \textsuperscript{12}C & 0.939 & \cite{weak11-12} \\
   5 & \textsuperscript{14}C $\to e^- + \nu+$ \textsuperscript{14}N & 0.730 & \cite{weak13-15} &
   6 & \textsuperscript{8}B $\to e^+ + \nu+$ 2\textsuperscript{4}He & 0.444 & \cite{weak5-10} \\
   7 & \textsuperscript{11}C $\to e^+ + \nu+$ \textsuperscript{11}B & 0.138 & \cite{weak11-12} &
   8 & \textsuperscript{12}N $\to e^+ + \nu+$ \textsuperscript{12}C & 0.140 & \cite{weak11-12} \\
   9 & \textsuperscript{13}N $\to e^+ + \nu+$ \textsuperscript{13}C & 0.0431  & \cite{weak13-15} &
   10 & \textsuperscript{14}O $\to e^+ + \nu+$ \textsuperscript{14}N & 0.0255 & \cite{weak13-15} \\
   11 & \textsuperscript{15}O $\to e^+ + \nu+$ \textsuperscript{15}N & 0.131 & \cite{weak13-15} &
   12 & H + $n$ $\to$ $\gamma$ + \textsuperscript{2}H & 7.00 & \cite{smith1993experimental}\\
   13 & \textsuperscript{2}H + $n$ $\to$ $\gamma$ + \textsuperscript{3}H & 15.0 & - &
   14 & \textsuperscript{3}He + $n$ $\to$ $\gamma$ + \textsuperscript{4}He & 15.0 & - \\
   15 & \textsuperscript{6}Li + $n$ $\to$ $\gamma$ + \textsuperscript{7}Li & 15.0 & - &
   16 & \textsuperscript{3}He + $n$ $\to$ $p$ + \textsuperscript{3}H & 10.0 & \cite{smith1993experimental} \\
   17 & \textsuperscript{7}Be + $n$ $\to$ $p$ + \textsuperscript{7}Li & 9.00 & \cite{smith1993experimental} &
   18 & \textsuperscript{6}Li + $n$ $\to$ $\alpha$ + \textsuperscript{3}H & 15.0 & - \\
   19 & \textsuperscript{7}Be + $n$ $\to$ $\alpha$ + \textsuperscript{4}He & 15.0 & - &
   20 & \textsuperscript{2}H + $p$ $\to$ $\gamma$ + \textsuperscript{3}He & 8.77 & \cite{NACREII} \\
   21 & \textsuperscript{3}H + $p$ $\to$ $\gamma$ + \textsuperscript{4}He & 15.0 & - &
   22 & \textsuperscript{6}Li + $p$ $\to$ $\gamma$ + \textsuperscript{7}Be & 40.0 & \cite{NACREII} \\
   23 & \textsuperscript{6}Li + $p$ $\to$ $\alpha$ + \textsuperscript{3}He & 14.0 & \cite{NACREII} &
   24 & \textsuperscript{7}Li + $p$ $\to$ $\alpha$ + \textsuperscript{4}He & 9.71 & \cite{NACREII} \\
   25 & \textsuperscript{2}H + $\alpha$ $\to$ $\gamma$ + \textsuperscript{6}Li & 22.0 & \cite{NACREII} &
   26 & \textsuperscript{3}H + $\alpha$ $\to$ $\gamma$ + \textsuperscript{7}Li & 10.1 & \cite{NACREII} \\
   27 & \textsuperscript{3}He + $\alpha$ $\to$ $\gamma$ + \textsuperscript{7}Be & 13.3 & \cite{NACREII} &
   28 & \textsuperscript{2}H + $D$ $\to$ $n$ + \textsuperscript{3}He & 2.28 & \cite{NACREII} \\
   29 & \textsuperscript{2}H + $D$ $\to$ $p$ + \textsuperscript{3}H & 0.477 & \cite{NACREII} &
   30 & \textsuperscript{3}H + $D$ $\to$ $n$ + \textsuperscript{4}He & 6.40 & \cite{NACREII} \\
   31 & \textsuperscript{3}He + $D$ $\to$ $p$ + \textsuperscript{4}He & 8.00 & \cite{smith1993experimental} &
   32 & \textsuperscript{3}He + \textsuperscript{3}He $\to$ 2$p$ + \textsuperscript{4}He & 7.89 & \cite{NACREII} \\
   33 & \textsuperscript{7}Li + $D$ $\to$ $n$ + $\alpha$ + \textsuperscript{4}He & 15.0 & - &
   34 & \textsuperscript{7}Be + $D$ $\to$ $p$ + $\alpha$ + \textsuperscript{4}He & 15.0 & - \\
   35 & \textsuperscript{3}He + \textsuperscript{3}H $\to$ $\gamma$ + \textsuperscript{6}Li & 15.0 & - &
   36 & \textsuperscript{6}Li + \textsuperscript{2}H $\to$ $n$ + \textsuperscript{7}Be & 15.0 & - \\
   37 & \textsuperscript{6}Li + \textsuperscript{2}H $\to$ $p$ + \textsuperscript{7}Li & 15.0 & - &
   38 & \textsuperscript{3}He + \textsuperscript{3}H $\to$ \textsuperscript{2}H + \textsuperscript{4}He  & 15.0 & - \\
   39 & \textsuperscript{3}H + \textsuperscript{3}H $\to$ 2$n$ + \textsuperscript{4}He & 15.0 & - &
   40 & \textsuperscript{3}H + \textsuperscript{3}H $\to$ $n$ + $p$ + \textsuperscript{4}He & 15.0 & - \\
   41 & \textsuperscript{7}Li + \textsuperscript{3}H $\to$ $n$ + \textsuperscript{9}Be & 15.0 & - &
   42 & \textsuperscript{7}Be + \textsuperscript{3}H $\to$ $p$  + \textsuperscript{9}Be & 15.0 & - \\
   43 & \textsuperscript{7}Li + \textsuperscript{3}He $\to$ $p$  + \textsuperscript{9}Be & 15.0 & - &
   44 & \textsuperscript{7}Li + $n$ $\to$ $\gamma$ + \textsuperscript{8}Li & 15.0 & - \\
   45 & \textsuperscript{10}B + $n$ $\to$ $\gamma$ + \textsuperscript{11}B & 15.0 & - &
   46 & \textsuperscript{11}B + $n$ $\to$ $\gamma$ + \textsuperscript{12}B & 15.0 & - \\
   47 & \textsuperscript{11}C + $n$ $\to$ $p$ + \textsuperscript{11}B & 15.0 & - &
   48 & \textsuperscript{10}B + $n$ $\to$ $\alpha$ + \textsuperscript{7}Li & 15.0 & - \\
   49 & \textsuperscript{7}Be + $p$ $\to$ $\gamma$ + \textsuperscript{8}B & 9.23 & \cite{NACREII} &
   50 & \textsuperscript{9}Be + $p$ $\to$ $\gamma$ + \textsuperscript{10}B & 11.4 & \cite{NACREII} \\
   51 & \textsuperscript{10}B + $p$ $\to$ $\gamma$ + \textsuperscript{11}C & 19.9 & \cite{NACREII} &
   52 & \textsuperscript{11}B + $p$ $\to$ $\gamma$ + \textsuperscript{12}C & 15.7 & \cite{NACREII} \\
   53 & \textsuperscript{11}C + $p$ $\to$ $\gamma$ + \textsuperscript{12}N & 15.0 & - &
   54 & \textsuperscript{12}B + $p$ $\to$ $n$ + \textsuperscript{12}C & 15.0 & - \\
   55 & \textsuperscript{9}Be + $p$ $\to$ $\alpha$ + \textsuperscript{6}Li & 16.6 & \cite{NACREII} &
   56 & \textsuperscript{10}B + $p$ $\to$ $\alpha$ + \textsuperscript{7}Be & 37.5 & \cite{NACREII} \\
   57 & \textsuperscript{12}B + $p$ $\to$ $\alpha$ + \textsuperscript{9}Be & 15.0 & - &
   58 & \textsuperscript{6}Li + $\alpha$ $\to$ $\gamma$ + \textsuperscript{10}B & 15.0 & - \\
   59 & \textsuperscript{7}Li + $\alpha$ $\to$ $\gamma$ + \textsuperscript{11}B & 26.9 & \cite{NACREII} &
   60 & \textsuperscript{7}Be + $\alpha$ $\to$ $\gamma$ + \textsuperscript{11}C & 43.0 & \cite{NACREII} \\
   61 & \textsuperscript{8}B + $\alpha$ $\to$ $p$ + \textsuperscript{11}C & 15.0 & - &
   62 & \textsuperscript{8}Li + $\alpha$ $\to$ $n$ + \textsuperscript{11}B & 15.0 & - \\
   63 & \textsuperscript{9}Be + $\alpha$ $\to$ $n$ + \textsuperscript{12}C & 21.4 & \cite{NACREII} &
   64 & \textsuperscript{9}Be + $D$ $\to$ $n$ + \textsuperscript{10}B & 15.0 & - \\
   65 & \textsuperscript{10}B + $D$ $\to$ $p$ + \textsuperscript{11}B & 15.0 & - &
   66 & \textsuperscript{11}B + $D$ $\to$ $n$ + \textsuperscript{12}C & 15.0 & - \\
   67 & \textsuperscript{4}He + $\alpha$ + $n$ $\to$ $\gamma$ + \textsuperscript{9}Be & 15.0 & - &
   68 & \textsuperscript{4}He + 2$\alpha$ $\to$ $\gamma$ + \textsuperscript{12}C & 15.0 & - \\
   69 & \textsuperscript{8}Li + $p$ $\to$ $n$ + $\alpha$ + \textsuperscript{4}He & 15.0 & - &
   70 & \textsuperscript{8}B + $n$ $\to$ $p$ + $\alpha$ + \textsuperscript{4}He & 15.0 & - \\
   71 & \textsuperscript{9}Be + $p$ $\to$ $D$ + $\alpha$ + \textsuperscript{4}He & 15.0 & - &
   72 & \textsuperscript{11}B + $p$ $\to$ 2$\alpha$ + \textsuperscript{4}He & 15.0 & - \\
   73 & \textsuperscript{11}C + $n$ $\to$ 2$\alpha$ + \textsuperscript{4}He & 15.0 & - &
   74 & \textsuperscript{12}C + $n$ $\to$ $\gamma$ + \textsuperscript{13}C & 15.0 & - \\
   75 & \textsuperscript{13}C + $n$ $\to$ $\gamma$ + \textsuperscript{14}C & 15.0 & - &
   76 & \textsuperscript{14}N + $n$ $\to$ $\gamma$ + \textsuperscript{15}N & 15.0 & - \\
   77 & \textsuperscript{13}N + $n$ $\to$ $p$ + \textsuperscript{13}C & 15.0 & - &
   78 & \textsuperscript{14}N + $n$ $\to$ $p$ + \textsuperscript{14}C & 15.0 & - \\
   79 & \textsuperscript{15}O + $n$ $\to$ $p$ + \textsuperscript{15}N & 15.0 & - &
   80 & \textsuperscript{15}O + $n$ $\to$ $\alpha$ + \textsuperscript{12}C & 15.0 & - \\
   81 & \textsuperscript{12}C + $p$ $\to$ $\gamma$ + \textsuperscript{13}N & 18.5 & \cite{NACREII} &
   82 & \textsuperscript{13}C + $p$ $\to$ $\gamma$ + \textsuperscript{14}N & 14.8 & \cite{NACREII} \\
   83 & \textsuperscript{14}C + $p$ $\to$ $\gamma$ + \textsuperscript{15}N & 15.0 & - &
   84 & \textsuperscript{13}N + $p$ $\to$ $\gamma$ + \textsuperscript{14}O & 22.4 & \cite{NACREII} \\
   85 & \textsuperscript{14}N + $p$ $\to$ $\gamma$ + \textsuperscript{15}O & 10.6 & \cite{NACREII} &
   86 & \textsuperscript{15}N + $p$ $\to$ $\gamma$ + \textsuperscript{16}O & 15.5 & \cite{NACREII} \\
   87 & \textsuperscript{15}N + $p$ $\to$ $\alpha$ + \textsuperscript{12}C & 83.7 & \cite{NACREII} &
   88 & \textsuperscript{12}C + $\alpha$ $\to$ $\gamma$ + \textsuperscript{16}O & 12.0 & \cite{NACREII} \\
   89 & \textsuperscript{10}B + $\alpha$ $\to$ $p$ + \textsuperscript{13}C & 15.0 & - &
   90 & \textsuperscript{11}B + $\alpha$ $\to$ $p$ + \textsuperscript{14}C & 15.0 & - \\
   91 & \textsuperscript{11}C + $\alpha$ $\to$ $p$ + \textsuperscript{14}N & 15.0 & - &
   92 & \textsuperscript{12}N + $\alpha$ $\to$ $p$ + \textsuperscript{15}O & 15.0 & - \\
   93 & \textsuperscript{13}N + $\alpha$ $\to$ $p$ + \textsuperscript{16}O & 15.0 & - &
   94 & \textsuperscript{10}B + $\alpha$ $\to$ $n$ + \textsuperscript{13}N & 15.0 & - \\
   95 & \textsuperscript{11}B + $\alpha$ $\to$ $n$ + \textsuperscript{14}N & 12.4 & \cite{NACREII} &
   96 & \textsuperscript{12}B + $\alpha$ $\to$ $n$ + \textsuperscript{15}N & 15.0 & - \\
   97 & \textsuperscript{13}C + $\alpha$ $\to$ $n$ + \textsuperscript{16}O & 26.5 & \cite{NACREII} &
   98 & \textsuperscript{11}B + \textsuperscript{2}H $\to$ $p$ + \textsuperscript{12}B & 15.0 & - \\
   99 & \textsuperscript{12}C + \textsuperscript{2}H $\to$ $p$ + \textsuperscript{13}C & 15.0 & - &
   100 & \textsuperscript{13}C + \textsuperscript{2}H $\to$ $p$ + \textsuperscript{14}C & 15.0 & - \\
  \hline
  \hline
\end{tabular}
\caption{Table of nuclear rates and associated uncertainties for reactions used in the BBN network. The list of reactions matches those in the latest version of the \texttt{AlterBBN} code~\cite{alterBBN,Arbey:2018zfh}.
References for tabulated uncertainties are provided, and those without a tabulated uncertainty were assigned a uniform error of 15\%. }
\label{tab:nuc_rates}
\end{table}


\newpage
\bibliographystyle{apsrev4-1}
\bibliography{bibliography}

\end{document}